\documentclass[twocolumn,twocolappendix]{aastex631}

\usepackage{amsmath}
\usepackage[T1]{fontenc}

\shorttitle{JVLA Observations of WL 17}
\shortauthors{Hashimoto et al.}

\DeclareUnicodeCharacter{02BC}{'}

\begin{document}

\title{JVLA Measurement of Grain Size in the Compact Dust Ring around Class I Protostar WL 17 }

\correspondingauthor{Jun Hashimoto}
\email{jhashimoto@asiaa.sinica.edu.tw}
\correspondingauthor{Hauyu Baobab Liu}
\email{baobabyoo@gmail.com}

\author[0000-0002-3053-3575]{Jun Hashimoto}
\affil{Academia Sinica Institute of Astronomy \& Astrophysics (ASIAA), 11F of Astronomy-Mathematics Building, AS/NTU, No.1, Sec. 4, Roosevelt Rd., Taipei 106319, Taiwan}
\affil{Astrobiology Center, National Institutes of Natural Sciences, 2-21-1 Osawa, Mitaka, Tokyo 181-8588, Japan}

\author[0000-0003-2300-2626]{Hauyu Baobab Liu}
\affiliation{Department of Physics, National Sun Yat-Sen University, No. 70, Lien-Hai Road, Kaohsiung City 80424, Taiwan, R.O.C.}
\affiliation{Center of Astronomy and Gravitation, National Taiwan Normal University, Taipei 116, Taiwan}

\author[0000-0001-9290-7846]{Ruobing Dong}
\affil{Kavli Institute for Astronomy and Astrophysics, Peking University, Beijing 100871, Peopleʼs Republic of China}
\affil{Department of Physics \& Astronomy, University of Victoria, Victoria, BC, V8P 5C2, Canada}

\author[0000-0001-5830-3619]{Beibei Liu}
\affil{Institute for Astronomy, School of Physics, Zhejiang University, 38 Zheda Road, Hangzhou 310027, China}

\author{Takayuki Muto}
\affil{Division of Liberal Arts, Kogakuin University, 1-24-2, Nishi-Shinjuku, Shinjuku-ku, Tokyo 163-8677, Japan}

\begin{abstract}
The maximum grain size in protoplanetary disks is a critical parameter for planet formation, as the efficiency of mechanisms like streaming instability and pebble accretion depend on grain size. Even young class 0/I objects, such as HL Tau, show substructures in their disks, indicating the potential for early planet formation. In this study, we investigated the grain size in the dust surrounding the class I object WL 17 using the Karl G. Jansky Very Large Array. Observations were conducted across seven frequency bands (Q, Ka, K, Ku, X, C, and S bands) ranging from 2 to 48 GHz, corresponding to wavelengths of 15 cm to 6.3 mm, with a spatial resolution exceeding 0\farcs5. While the ring structure at 0\farcs1 of WL 17 remains unresolved in our data, its emission is clearly detected at all observed frequencies, except at 2 GHz. To estimate the maximum grain size ($a_{\rm max}$) within the ring, we compared the observed spectral energy distribution (SED) with theoretical SEDs calculated for various $a_{\rm max}$ values using radiative transfer models. Assuming the dust opacity follows the DSHARP model, our analysis suggests that certain structures internal to the ring achieved a maximum grain size of approximately 4.2 mm. Additionally, we discuss the gravitational stability of the ring and the potential planetary core mass that could form through pebble accretion within the structure.
\end{abstract}


\section{Introduction}

The discovery of substructures in the HL Tau disk \citep{almapart+15} represents a significant breakthrough in the field of planet formation. Traditionally, planet formation in the standard core-accretion scenario has been believed to take more than 1 Myr \citep[e.g.,][]{poll1996}. However, the presence of substructures in the young HL Tau disk, which is less than 1 Myr, suggests that planet formation occurs much earlier than previously expected, potentially indicating the presence of forming planets in the disk \citep{Dong2015gap}. The presence of giant planets is often inferred by assuming that most substructures in disks are formed by these planets \citep[e.g.,][]{Paardekooper2023PPVII}. Furthermore, recent studies of the atmospheres in directly imaged exoplanets suggest that a solid accretion mass of approximately 100 $M_\earth$ is needed to explain the metal enrichment in these exoplanets \citep[e.g., $\beta$ Pic b;][]{GRAVITY2020betaPicb}. Given that typical protoplanetary disks around Class II objects with a solar mass have a solid mass of approximately 10 $M_\earth$ inferred by assuming optically thin (sub)millimeter emission \citep[e.g.,][]{ansdell2017a}, much less than the total solid mass in typical planetary systems \citep{Manara2018Mass}, planet formation earlier than 1 Myr might be necessary to unravel this puzzle.

Nevertheless, recent findings from the eDISK project, The Atacama Large Millimeter/submillimeter Array (ALMA) large program in cycle 7, which focuses on searching for substructures around protostars younger than 1 Myr, reported only a few distinct substructures \citep{Ohashi2023edisk}. This contrasts with class II objects, where substructures have commonly been observed in disks \citep[e.g., DSHARP in][]{Andrews2018DSHARP}. The scarcity of substructures around protostars suggests that they may be rare in these early stages of planet formation. The lack of substructure may also be attributed to the disk's optical depth in ALMA wavelengths at the Class 0/I stage \citep[e.g.,][]{Li2017ApJ...840...72L,Galvan2018ApJ...868...39G,Ko2020ApJ...889..172K,Liu2021ApJ...914...25L,Xu2022ApJ...934..156X}. This gives additional motivation to explore longer, centimeter wavelengths to determine the amount of material in the disk. By studying the physical and chemical properties of substructures around protostars, we could gain a better understanding of the rapid planet formation processes that occur within a timescale of less than 1 Myr.

The maximum grain size ($a_{\rm max}$) is one of the key parameters in planet formation. The efficiency of mechanisms such as streaming instability \citep[e.g.,][]{joha2014a} and pebble accretion \citep[e.g.,][]{Ormel2010Pebble,Lambrechts2012Pebble}, key steps in planet formation, crucially depend on $a_{\rm max}$ \citep[e.g.,][]{Liu2020,Drazkowska2023PPVII}. 

Two methods exist for determining the sizes of dust particles based on millimeter/centimeter (mm/cm) continuum emission. The first approach involves observing polarized dust continuum emission in the mm/cm range, resulting from scattering \citep[e.g.,][]{Kataoka2015Polarization}. The polarization fraction in the dust emission reaches its maximum when the dust grains grow to a size of approximately $a_{\rm max} \sim \lambda/2\pi$, where $\lambda$ represents the observing wavelength. This method requires a signal-to-noise ratio (SNR) of approximately 100--1,000 in Stokes~$I$ and is well-suited for studying bright objects like DSHARP disks (Andrews et al.\ 2018). The second approach involves constructing the spectral energy distribution \citep[SED, e.g.,][]{Liu2019SED}. This method relies on the fact that grains emit thermal radiation most efficiently at wavelengths similar to their sizes \citep[e.g.,][]{Draine2006}. By analyzing the SED, information about the grain sizes can be inferred.

The grain sizes in disks around class 0/I objects have been estimated to be sub-millimeter sizes by ALMA polarization observations \citep[e.g., HL Tau, HH 212, TMC-1A, GSS 30 IRS 1,][]{Stephens2017HLTau,Lee2021HH212,Aso2021TMC-1A,Sadavoy2019Polarization}. ALMA observations are particularly sensitive to sub-millimeter-sized grains, whereas the Karl G. Jansky Very Large Array (JVLA) is more suited for detecting millimeter- to centimeter-sized grains. This makes JVLA observations advantageous for investigating grain growth at these larger scales. Prescence of grain growth to centimeter sizes has been suggested in some disks around protostars, such as CB 26 and EC 53 \citep[e.g.,][]{Zhang2021CB26,Lee2020EC53}.

Several class 0/I objects, such as HL~Tau, WL~17, GY~91, L1527, Oph~IRS~63, HH111~VLA1, TMC-1A, IRAS~04169$+$2702, L1489~IRS, Oph~A~SM1, have displayed substructures in their disks \citep{almapart+15,Sheehan2017WL17,Sheehan2018GY91,Nakatani2020ApJ...895L...2N,Segura-Cox2020OphIRS63,Lee2020HH111VLA1,Aso2021TMC-1A,Ohashi2023edisk,Maureira2024A&A...689L...5M}. \citet{Jiang2023Pebble} propose that dust rings serve as highly efficient regions for planet formation through pebble accretion. In this paper, we present comprehensive multi-wavelength observations of WL 17 conducted with the JVLA.

WL 17 is an M3-type class-I protostar \citep{McClure2010Oph} with a mass of 0.3 $M_\sun$ \citep{Han2023WL17}, situated at a distance of 135.6 pc in the Ophiuchus molecular cloud \citep{gaia2016,gaia2022}, and it is known to possess a dust ring with a low inclination angle of 34 deg \citep{Sheehan2017WL17,Shoshi2024WL17} while being deeply embedded in its envelope \citep{evan+2009}. \citet{Sadavoy2019Polarization} found that WL 17 showed no polarization signal at $\lambda=1.3$ mm, with a 3 $\sigma$ upper limit of $< 0.3$ \%. This suggests that the grain size in WL 17's ring is either smaller or larger than $a_{\rm max} \sim \lambda/2\pi \sim 200$ $\mu$m. To further determine the grain size in WL 17, analyzing the SED including ALMA and JVLA data is helpful. 

\section{Observations and results} \label{sec:obs}

The JVLA observations of WL 17, spanning a frequency range of 2--48 GHz with the C configuration, were carried out under program ID 23A-124 (PI: J. Hashimoto). Detailed information about the observations is provided in Table \ref{tab:obs}. The data were collected over five epochs. Given that WL~17's dust ring at $r \sim 0\farcs1$ \citep{Sheehan2017WL17,Shoshi2024WL17} cannot be spatially resolved from free-free or synchrotron emissions near the central star with the C configuration's beam size of $\gtrsim$0\farcs5, lower-frequency observations (2--12 GHz) were performed during all five epochs to estimate the contribution from these emission mechanisms. The observation dates for each frequency band are summarized in Table~\ref{tab:image}.

\begin{deluxetable*}{ll}[htb!]
\tablewidth{0pt} 
\tablecaption{JVLA observations with C configuration \label{tab:obs}}
\tablehead{
\colhead{Observations}      & \colhead{}    
}
\startdata
Observing date (UT)          & Jan.27 (Ep.1), Feb.5 (Ep.2), Feb. 18 (Ep.3), Feb. 20 (Ep.4), Feb. 21 (Ep.5), 2024 \\
Project code                 & 24A-001 (PI: J. Hashimoto)       \\
Central frequency (GHz)      & 44 (Q), 33 (Ka), 20 (K), 15 (Ku), 10 (X), 6 (C), 3 (S) \\
Continuum band width (GHz)   & 8 (Q, Ka, K), 6 (Ku), 4 (X, C), 2 (S) \\
Total time on source (min)   & 7.4 (Q), 7.3 (Ka), 35.5 (K), 70.2 (Ku), 34.6 (X), 37.0 (C), 34.6 (S) \\
Number of antennas           & 27 (Jan.27, Feb.5, 18, 21), 26 (Feb.20) \\
Baseline lengths (km)        & 0.0448 to 3.4                  \\
Bandpass and Flux calibrator & 3C286                     \\
Phase calibrator             & J1625$-$2527, J1626$-$2951 \\
\enddata
\end{deluxetable*}

The data were calibrated using the Common Astronomy Software Applications (CASA) package \citep{Bean2022PASP..134k4501C}, following the calibration pipeline provided by JVLA. As self-calibration of the visibilities did not significantly improve the SNR, we did not apply self-calibration. The multi-term multi-frequency synthesis imaging with {\tt nterm}~=~2 \citep{Rau2011MultiscaleClean} and the robust clean parameter of 2 was performed using the CASA {\tt tclean} task, as summarized in Table~\ref{tab:image}. 

\begin{figure*}[htbp]
    \centering
    \includegraphics[width=0.95\linewidth]{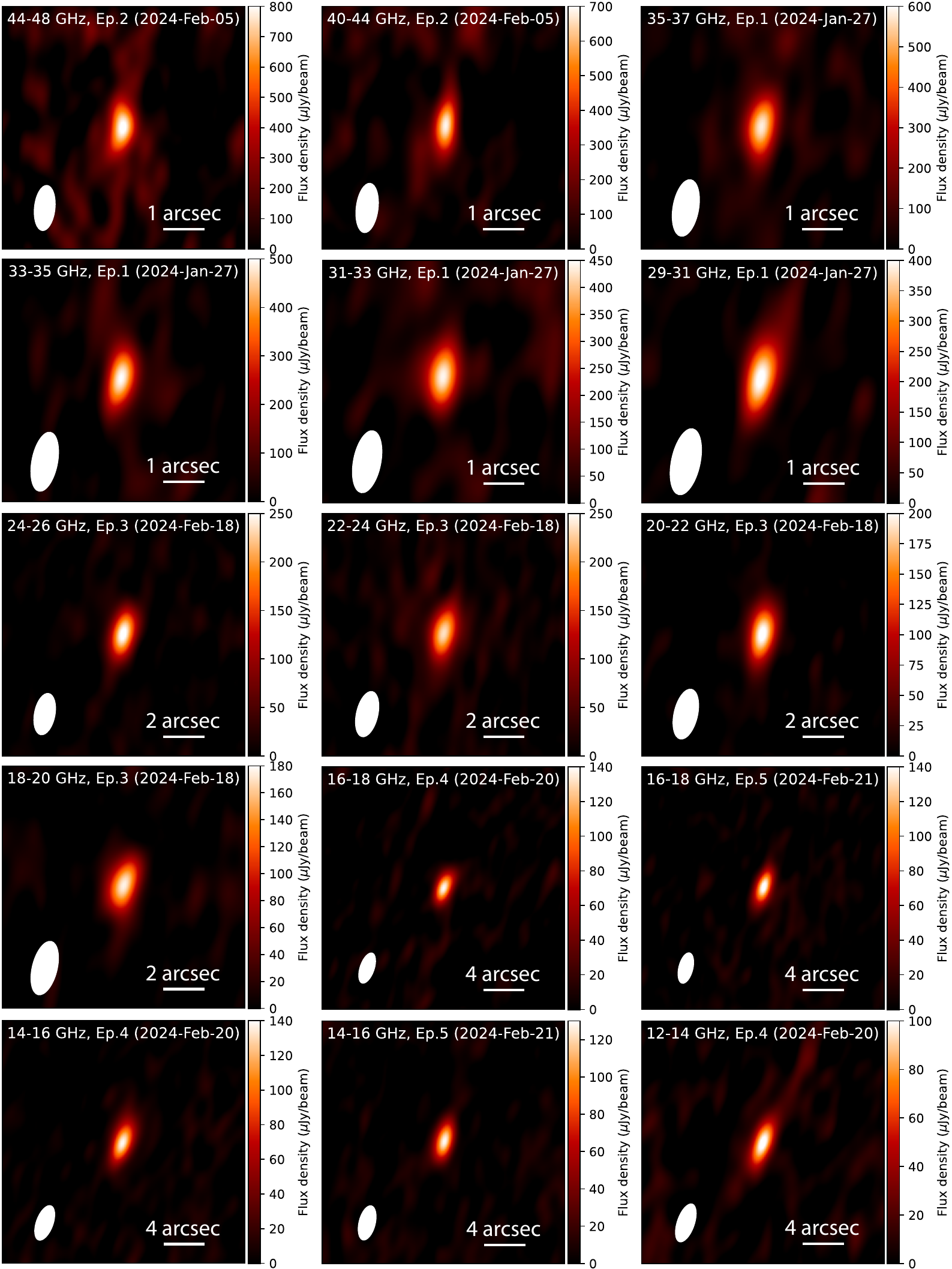} 
    \caption{
    The JVLA intra-band images of WL 17. The observing frequency and epoch are labeled in each panel. The center of the image corresponds to the stellar position in FK5 J2000.0 at the observing epoch of 2024.1, calculated as (16$^{\rm h}$27$^{\rm m}$06$\fs$7639, $-24^{\circ}$38$^{\rm m}$15$\fs$53589) (see the text for details). The offset between the stellar position and the peak emission is approximately 0\farcs1--0\farcs2. The synthesized beams are represented as white ellipses in the lower-left corner of each image. The beam shape and r.m.s. noise are summarized in Table \ref{tab:image}.
    }
    \label{fig:cband-image1}
\end{figure*}

\begin{figure*}[htbp]
    \includegraphics[width=\linewidth]{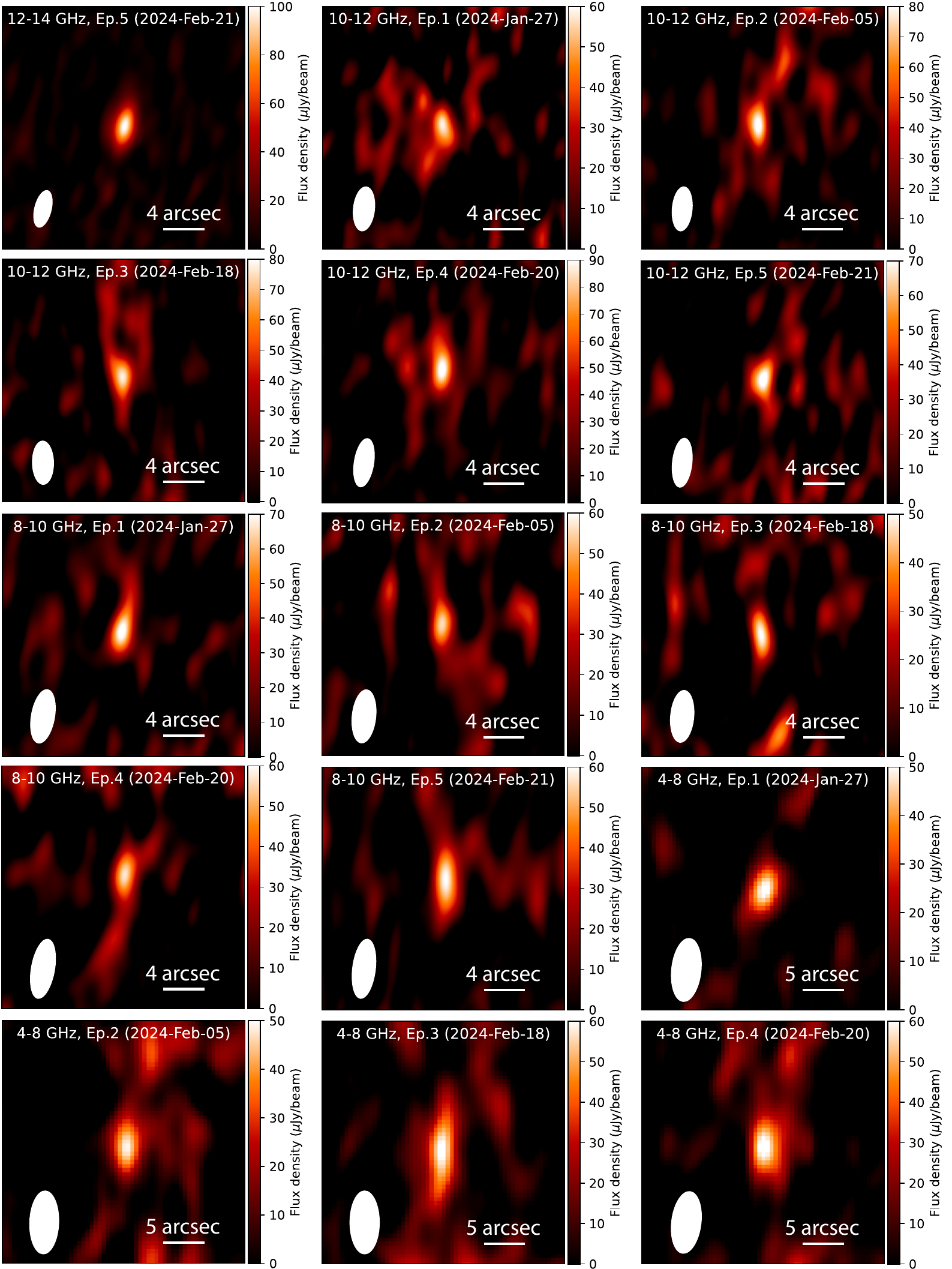} 
    \caption{
    Same as Figure \ref{fig:cband-image1}, except different intra-band frequency and epoch.
    }
    \label{fig:cband-image2}
\end{figure*}

\begin{figure*}[htbp]
    \includegraphics[width=\linewidth]{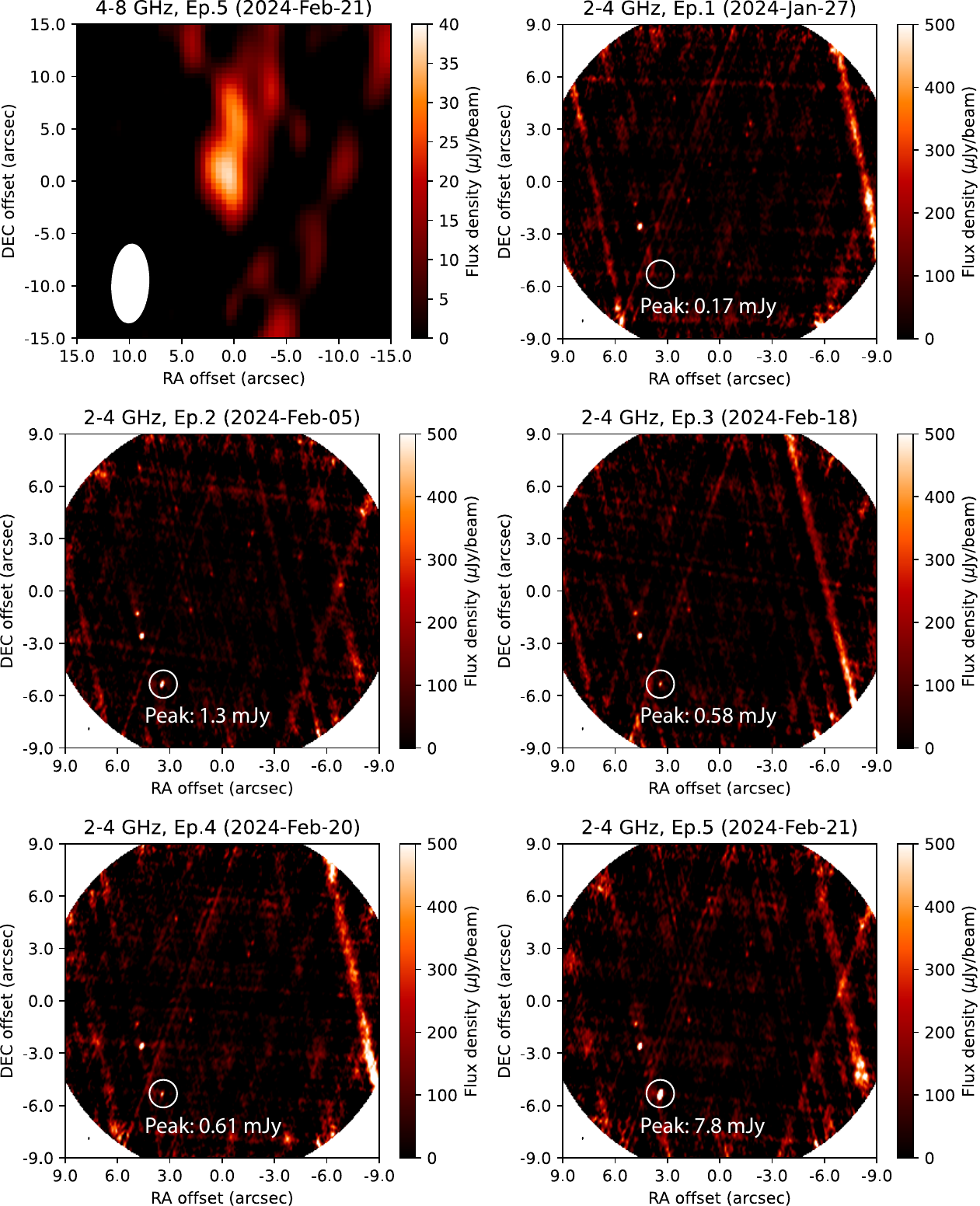} 
    \caption{
    Same as Figure \ref{fig:cband-image1}, except different intra-band frequency and epoch. The white circles highlight a radio burst from [GY92] 253 (see text in \S \ref{sec:obs}).
    }
    \label{fig:cband-image6}
\end{figure*}

Figures \ref{fig:cband-image1}--\ref{fig:cband-image6} present JVLA images of WL 17 observed across the 2–48 GHz frequency range. The center of the image corresponds to the stellar position in FK5 J2000.0 at the observing epoch of 2024.1, calculated as (16$^{\rm h}$27$^{\rm m}$06$\fs$7639, $-24^{\circ}$38$^{\rm m}$15$\fs$53589) using the proper motion of (pmRA, pmDEC) = ($-10.0$, $-27.9$) mas yr$^{-1}$ \citep{Ducourant2017Oph-PM}. To estimate the grain size by analyzing the spectral energy distribution (SED) with the highest possible number of data points, the observed data were divided into intra-band segments with 4 GHz or 2 GHz bandwidths, except for the 2--8 GHz range due to its lower signal-to-noise ratio (SNR). Significant signals were detected at 4–48 GHz, as shown in Figures \ref{fig:cband-image1}--\ref{fig:cband-image6} and summarized in Table \ref{tab:image}. 

Since the emission from WL 17 is not spatially resolved at the spatial resolution of $\gtrsim0\farcs5$, we use the CASA \texttt{uvmodelfit} task with a component model type of `Point source' (\texttt{comptype=`P'}) to estimate the flux density in the visibility domain. Prior to applying the \texttt{uvmodelfit} task, we subtract bright nearby sources using the CASA \texttt{uvsub} task. The model visibilities used for this subtraction are generated with the CASA \texttt{tclean} task. As an alternative measurement, we also use the peak flux density of WL 17 from Figures \ref{fig:cband-image1}--\ref{fig:cband-image6} as a proxy for the total flux density. Both methods yield similar flux densities to within 10\%, as shown in Table \ref{tab:image}. For the subsequent SED analysis, we adopt the flux density estimated with \texttt{uvmodelfit}, as the peak flux density may be affected by image r.m.s.\ noise. 

For the lower-frequency observations at 2--4 GHz shown in Figure \ref{fig:cband-image6}, significant sidelobes caused by strong radio emissions outside the field of view (FoV) hinder accurate flux density measurements for WL 17. Interestingly, [GY92] 253, marked by white circles in Figure \ref{fig:cband-image6}, exhibited a radio burst on 2024-Feb-21. Its brightness increased to 7.8 mJy---approximately an order of magnitude higher than in other epochs, where it ranged from 0.17 to 1.3 mJy. Such radio bursts from young stellar objects are often associated with magnetic activity \citep[e.g.,][]{Gudel2002ARA&A..40..217G}. 

\begin{deluxetable*}{ccccccc}[htbp]
\tablewidth{0pt} 
\tablecaption{Intra-band flux density and imaging parameters\label{tab:image}}
\tablehead{
\colhead{Intra-band frequency} & \colhead{Flux density} & \colhead{Flux density} & \colhead{Adapted error} & \colhead{Beam shape} & \colhead{r.m.s. noise}   & \colhead{Epoch\tablenotemark{c}}\\
\colhead{}                     & \colhead{---{\tt uvmodelfit}---}& \colhead{---peak flux---} & \colhead{in flux density\tablenotemark{a}} & \colhead{}           & \colhead{in image} & \colhead{} \\
\colhead{(GHz)}      & \colhead{($\mu$Jy)} & \colhead{($\mu$Jy)} & \colhead{($\mu$Jy)} & \colhead{}           & \colhead{($\mu$Jy/beam)} & \colhead{} 
}
\startdata
44-48 & 847.2 & 830.2 & 127.1 & $1\farcs08\times0\farcs50$ & 76.1 & Ep.2 \\
40-44 & 676.7 & 692.2 & 101.5 & $1\farcs18\times0\farcs53$ & 50.5 & Ep.2 \\ 
35-37 & 578.3 & 579.8 &  86.7 & $1\farcs36\times0\farcs62$ & 36.4 & Ep.1 \\
33-35 & 501.9 & 502.6 &  75.3 & $1\farcs42\times0\farcs62$ & 31.5 & Ep.1 \\
31-33 & 442.9 & 442.9 &  66.4 & $1\farcs49\times0\farcs66$ & 30.8 & Ep.1 \\
29-31 & 411.9 & 413.4 &  61.8 & $1\farcs59\times0\farcs69$ & 30.0 & Ep.1 \\
24-26 & 259.9 & 260.9 &  39.0 & $2\farcs01\times1\farcs01$ & 11.5 & Ep.3 \\
22-24 & 237.8 & 235.8 &  35.4 & $2\farcs20\times1\farcs03$ & 13.2 & Ep.3 \\
20-22 & 210.4 & 209.6 &  31.6 & $2\farcs42\times1\farcs13$ & 12.8 & Ep.3 \\ 
18-20 & 175.2 & 176.1 &  26.3 & $2\farcs59\times1\farcs20$ & 10.0 & Ep.3 \\
16-18 & 139.9 & 140.4 &   7.2 & $3\farcs05\times1\farcs41$ &  7.2 & Ep.4 \\
16-18 & 146.0 & 146.4 &   7.3 & $2\farcs99\times1\farcs39$ &  6.7 & Ep.5 \\
14-16 & 139.1 & 138.8 &   7.0 & $3\farcs48\times1\farcs62$ &  7.0 & Ep.4 \\
14-16 & 129.2 & 129.4 &   6.5 & $3\farcs36\times1\farcs59$ &  6.3 & Ep.5 \\
12-14 & 107.9 & 106.5 &   7.2 & $3\farcs80\times1\farcs69$ &  7.2 & Ep.4 \\
12-14 & 106.0 & 101.2 &   6.4 & $3\farcs72\times1\farcs68$ &  6.4 & Ep.5 \\
10-12 &  63.5 &  58.7 &  14.1 & $4\farcs39\times2\farcs13$ & 14.1 & Ep.1 \\
10-12 &  82.2 &  84.4 &  13.5 & $4\farcs38\times2\farcs02$ & 13.5 & Ep.2 \\
10-12 &  72.9 &  78.0 &  14.7 & $4\farcs28\times2\farcs12$ & 14.7 & Ep.3 \\
10-12 &  93.7 &  94.3 &  15.0 & $4\farcs82\times1\farcs98$ & 15.0 & Ep.4 \\
10-12 &  73.6 &  74.8 &  15.1 & $4\farcs84\times1\farcs95$ & 15.1 & Ep.5 \\
8-10  &  74.1 &  73.5 &  11.5 & $5\farcs33\times2\farcs34$ & 11.5 & Ep.1 \\
8-10  &  57.1 &  56.3 &  13.9 & $5\farcs23\times2\farcs36$ & 13.9 & Ep.2 \\
8-10  &  50.0 &  50.5 &  12.8 & $5\farcs14\times2\farcs33$ & 12.8 & Ep.3 \\
8-10  &  56.2 &  57.4 &  12.8 & $5\farcs87\times2\farcs32$ & 12.8 & Ep.4 \\
8-10  &  63.6 &  60.6 &  12.8 & $5\farcs86\times2\farcs28$ & 12.8 & Ep.5 \\
4-8   &  55.4 &  51.6 &  10.3 & $7\farcs65\times3\farcs59$ & 10.3 & Ep.1 \\
4-8   &  52.0 &  49.0 &  10.0 & $7\farcs62\times3\farcs56$ & 10.0 & Ep.2 \\
4-8   &  61.7 &  61.3 &  10.0 & $7\farcs58\times3\farcs58$ & 10.0 & Ep.3 \\
4-8   &  67.3 &  64.2 &   9.8 & $7\farcs51\times3\farcs61$ &  9.8 & Ep.4 \\
4-8   &  36.8 &  37.3 &  10.5 & $7\farcs58\times3\farcs59$ & 10.5 & Ep.5 \\
2-4   & non detection & non detection &  32.5 & $14\farcs1\times6\farcs58$ & 32.5 & Ep.1 \\
2-4   & non detection & non detection &  29.7 & $14\farcs0\times6\farcs51$ & 29.7 & Ep.2 \\
2-4   & non detection & non detection &  32.4 & $13\farcs9\times6\farcs47$ & 32.4 & Ep.3 \\
2-4   & non detection & non detection &  36.0 & $13\farcs8\times6\farcs52$ & 36.0 & Ep.4 \\
2-4   & non detection & non detection &  37.8 & $16\farcs2\times6\farcs34$ & 37.8 & Ep.5 \\
\hline
\enddata
\tablenotetext{a}{
The absolute flux density accuracy of the JVLA is 5\% for the L- through Ku-bands and 15\% for the three higher-frequency bands (\url{https://science.nrao.edu/facilities/vla/docs/manuals/oss2024A}). We conservatively adopt the larger of the absolute flux uncertainty or the image r.m.s.\ noise as the error in flux density when making comparisons. The errors estimated by the \texttt{uvmodelfit} task are typically about 50\% of the image r.m.s.\ noise.
}\vspace{-0.2cm}
\tablenotetext{b}{Detailed observing date is provided in Table \ref{tab:obs}.}\vspace{-0.2cm}
\end{deluxetable*}

\begin{figure*}[htb!]
\centering
\begin{tabular}{cc}
\includegraphics[width=8cm]{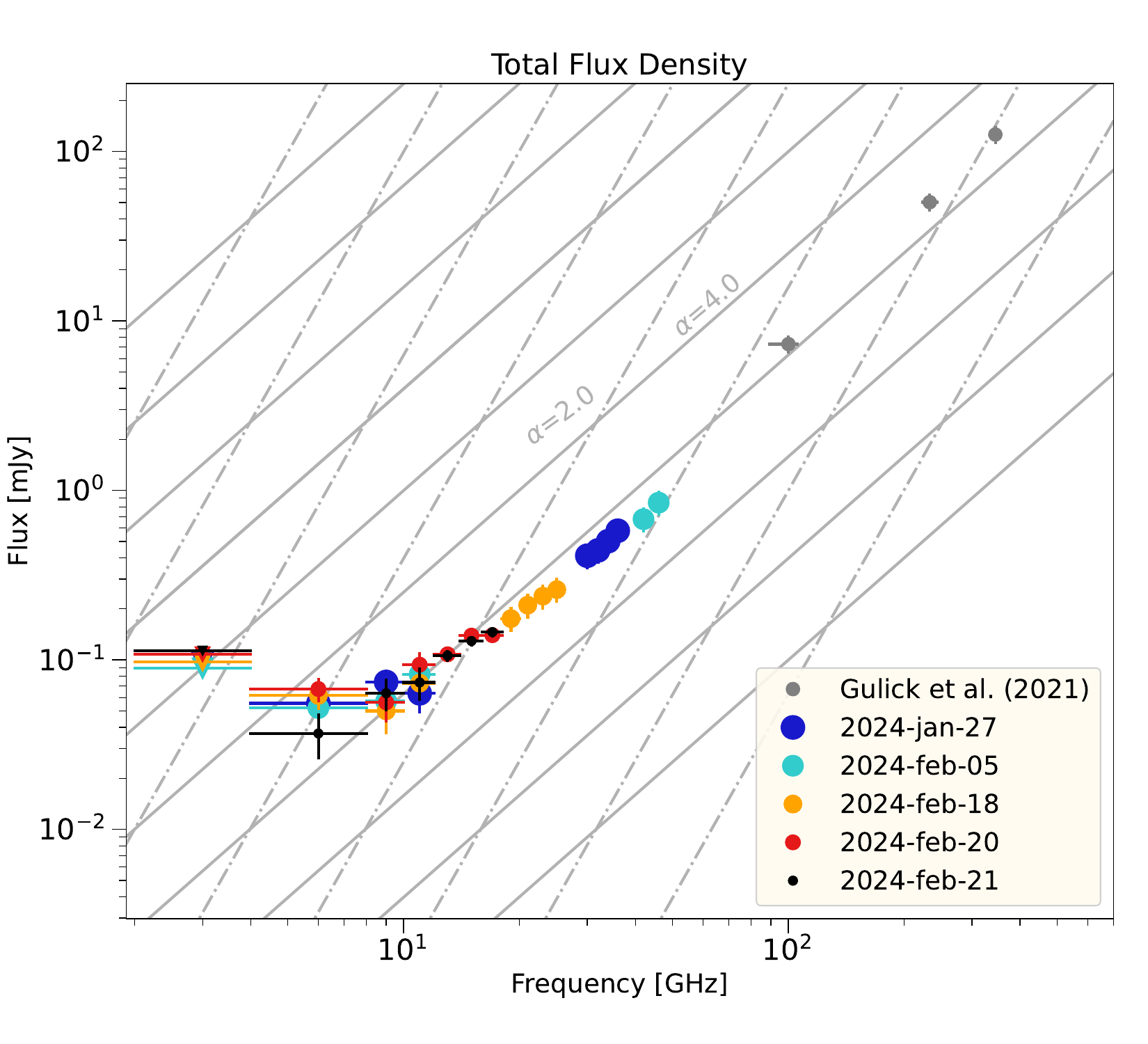} &
\includegraphics[width=8cm]{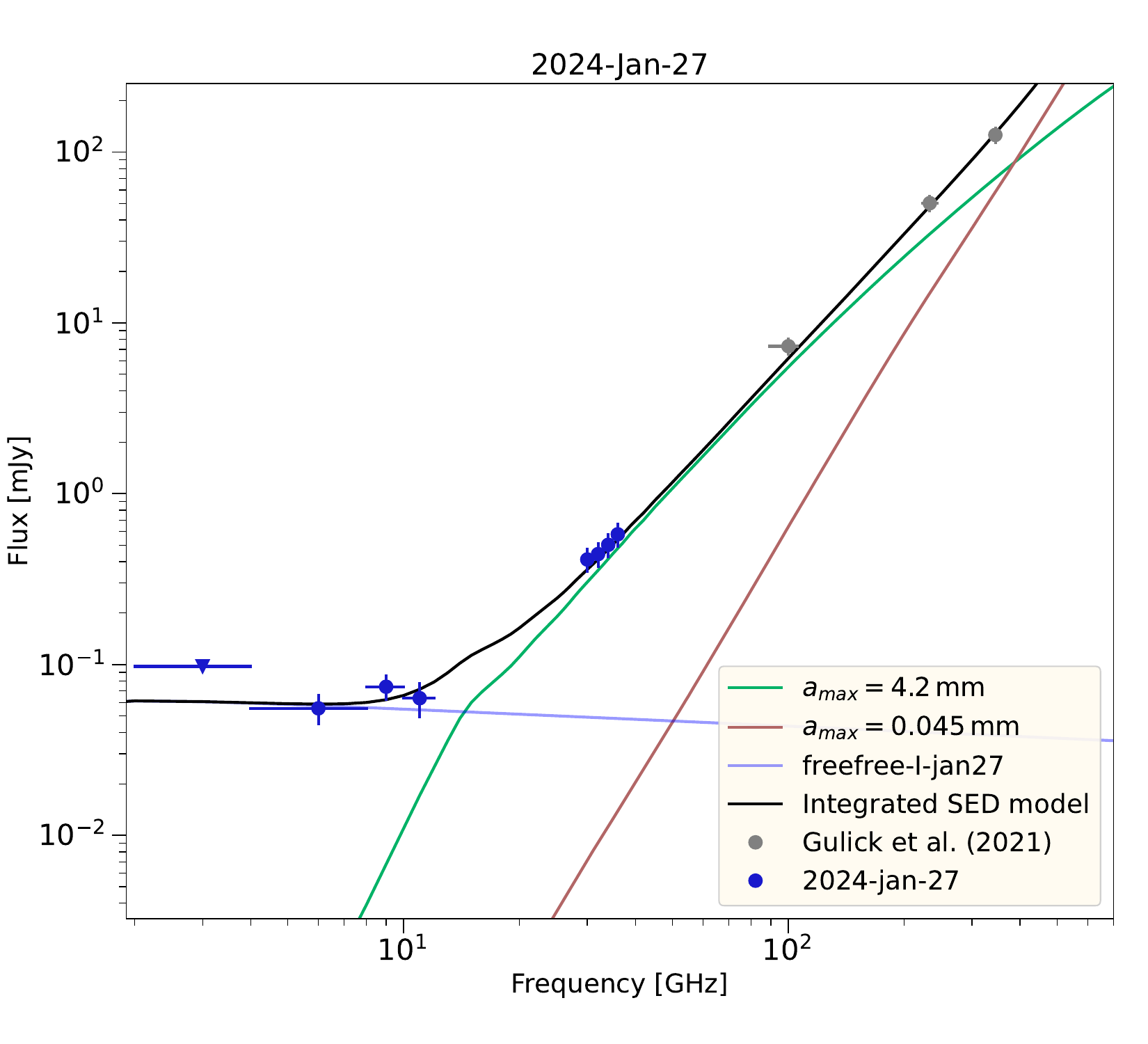} \\
\includegraphics[width=8cm]{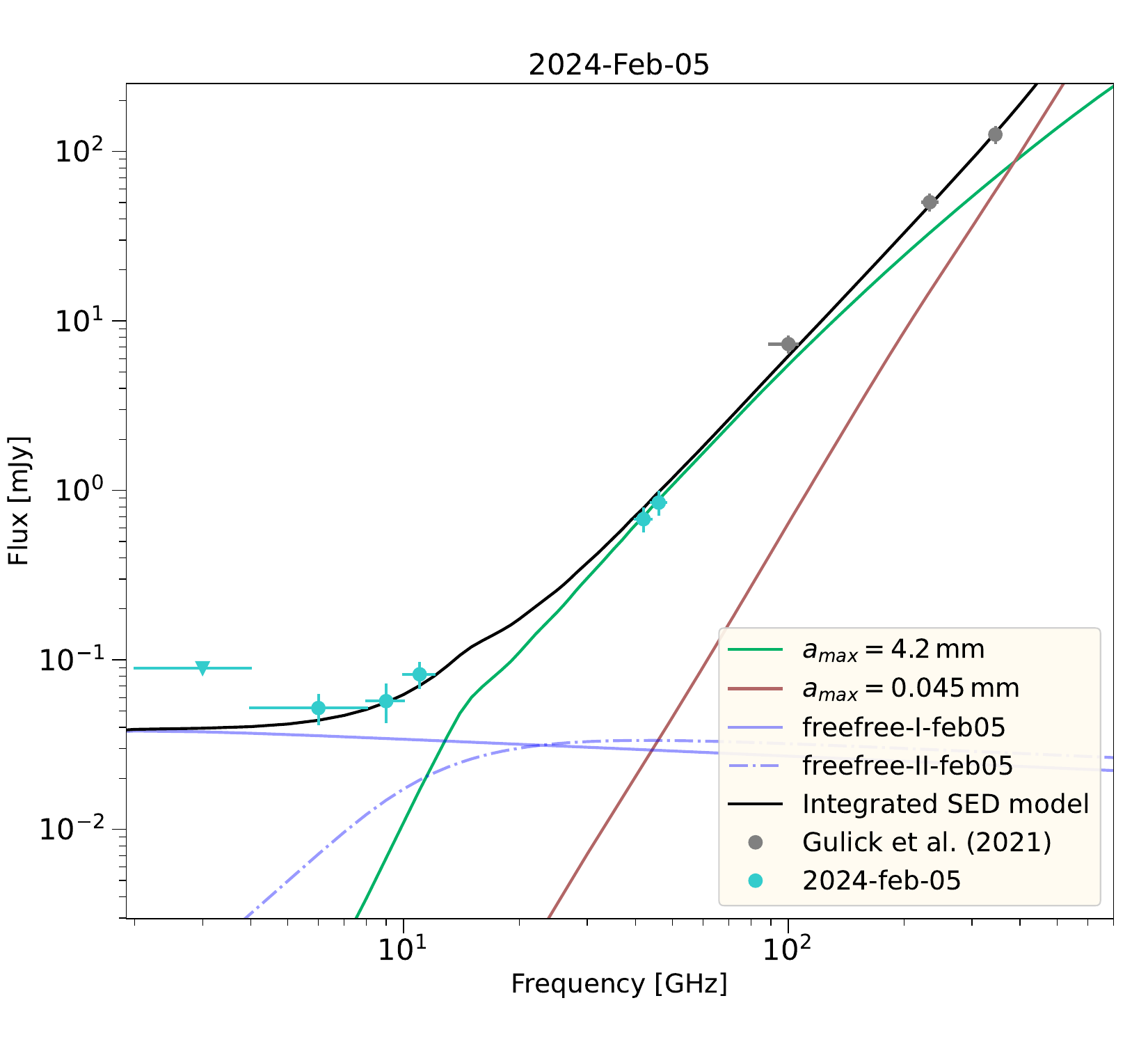} &
\includegraphics[width=8cm]{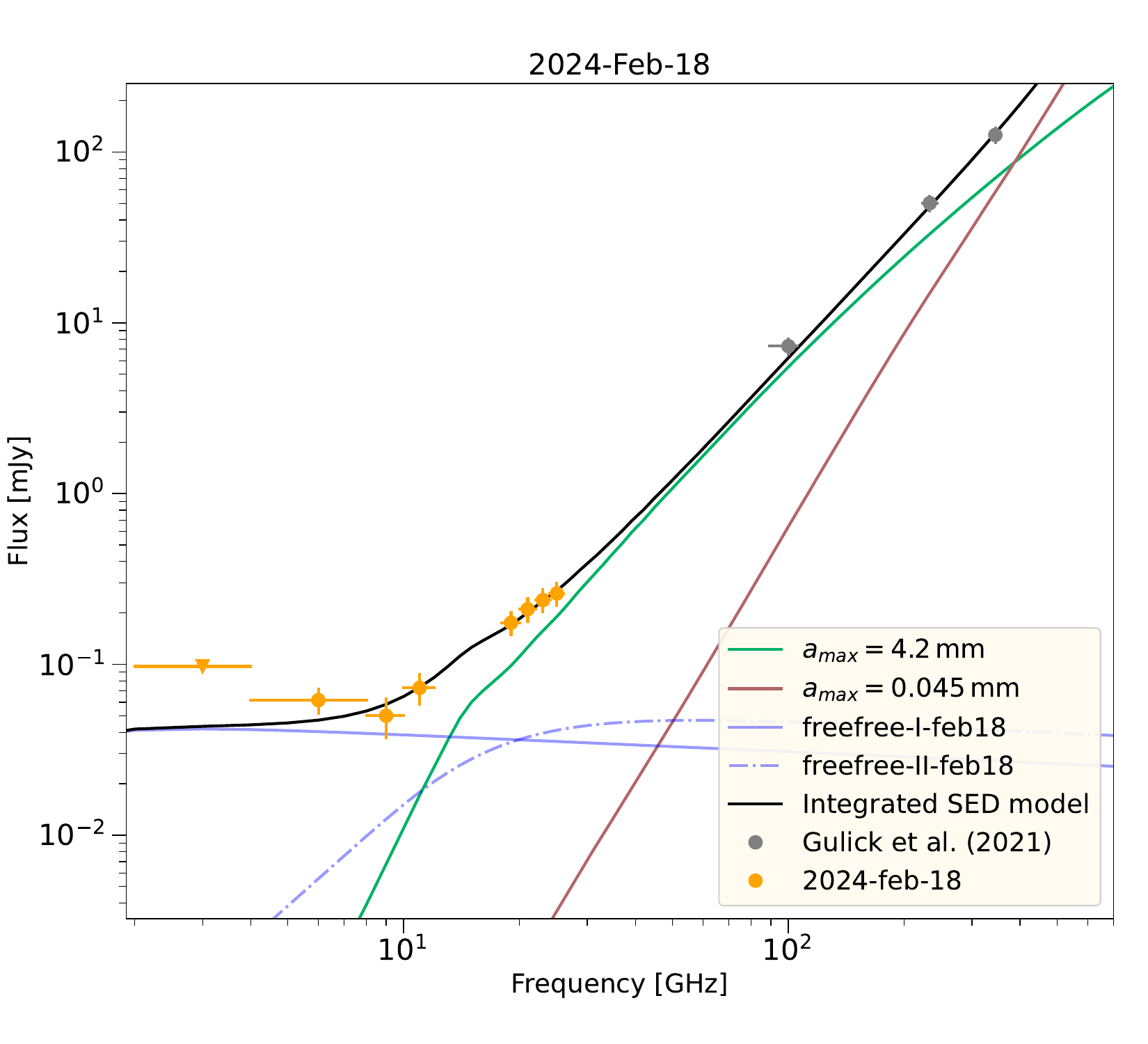} \\
\includegraphics[width=8cm]{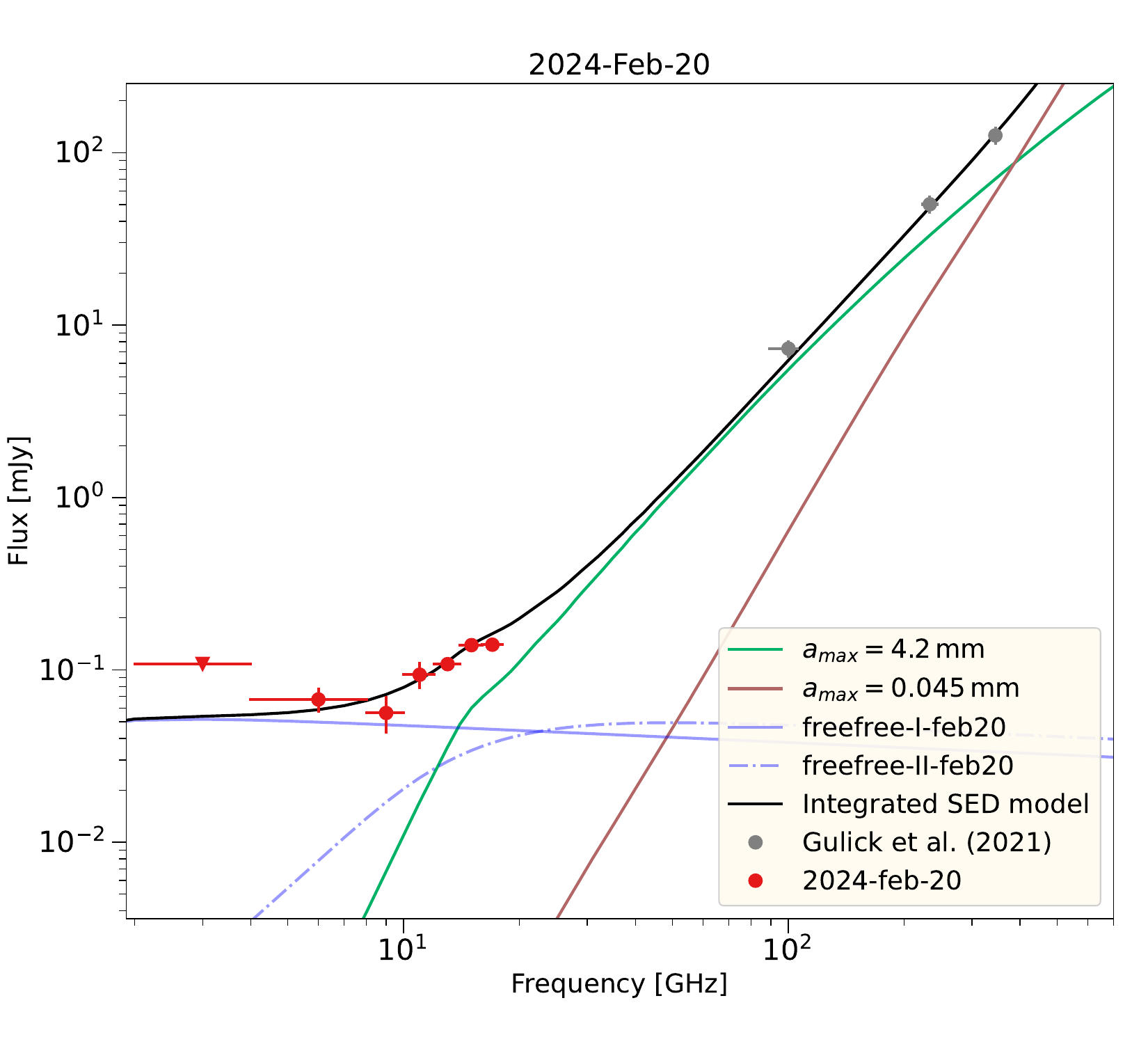} &
\includegraphics[width=8cm]{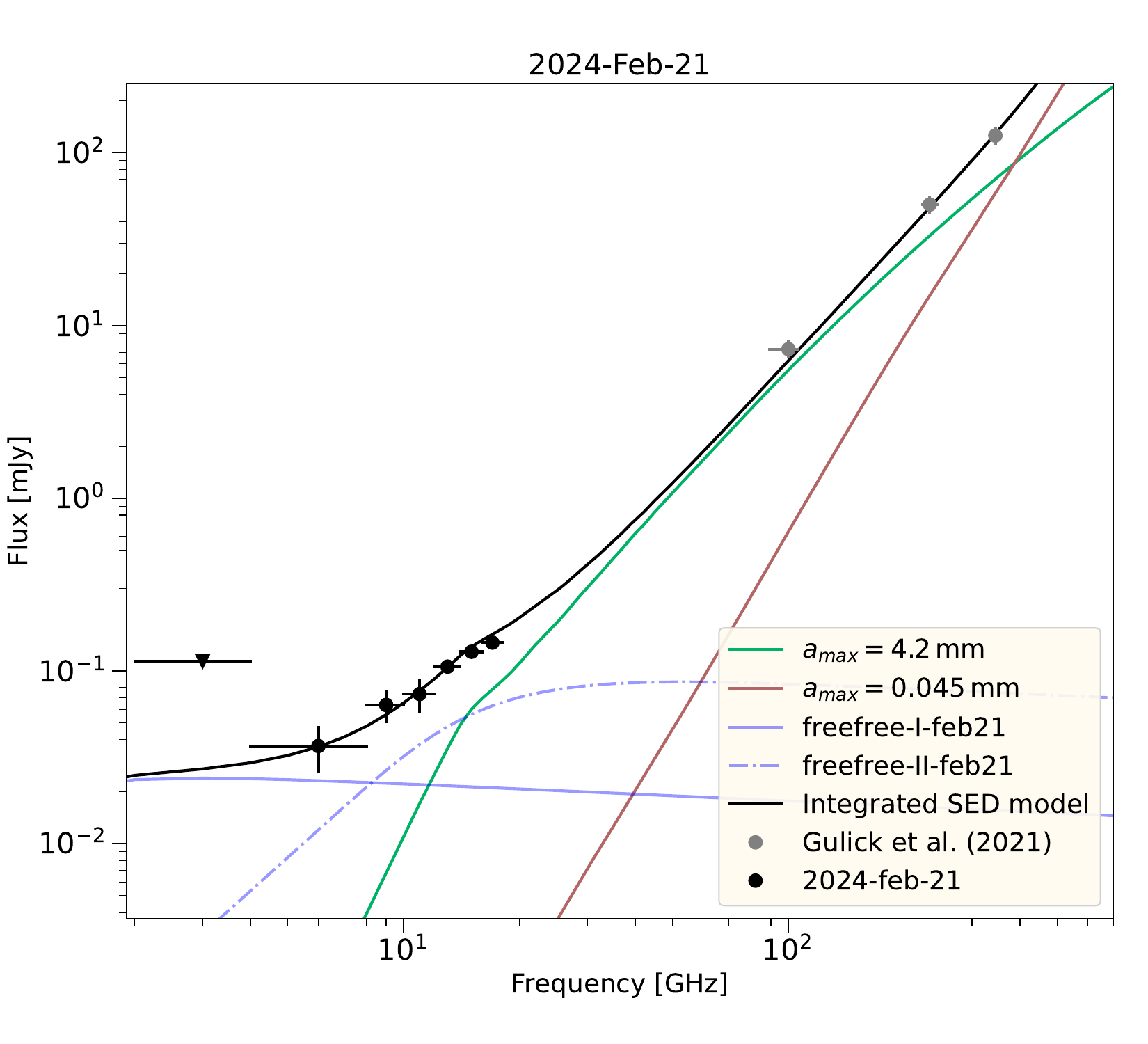} \\
\end{tabular}
\caption{
    The 2--345 GHz spectral profile of WL~17. 
    Top left panel shows the flux densities obtained from all of our JVLA observations (Table~\ref{tab:image}) and previous ALMA observations \citep{Gulick2021WL17}, assuming a nominal 10\% (1$\sigma$) absolute flux uncertainty. The ALMA fluxes from \citet{Gulick2021WL17} are not direct measurements but model-derived (their Table 3), and may be underestimated. Some symbols are larger than their error bars. Gray solid and dash-dotted lines indicate spectral indices of 2.0 and 4.0, respectively. Each remaining panel shows fluxes from a specific JVLA epoch overlaid with best-fit dust and free-free emission models.
}
\label{fig:sed}
\end{figure*}

The SEDs taken from our observations are summarized in Figure \ref{fig:sed}, which are compared with the previous ALMA observations published in \citet{Gulick2021WL17}.
At $>$90 GHz frequencies, the spectral index ($\alpha$) is slightly larger than 2.0 as summarized in Table \ref{tab:spindex}.
Meanwhile, the spectral indices are comparably smaller than 2.0 at 14--48 GHz.
At $<$14 GHz, the spectral indices are considerably lower than 2.0.
Both flux densities and spectral indices present non-negligible time variation.
Our interpretation for these observations are given in the following Section \ref{sec:discussion}.

\begin{deluxetable}{ccc}[htbp]
\tablewidth{0pt} 
\tablecaption{Spectral index ($\alpha$) \label{tab:spindex}}
\tablehead{
\colhead{Frequency range} & \colhead{Spectral index} & \colhead{Epoch\tablenotemark{a} or} \\
\colhead{(GHz)}           & \colhead{}               & \colhead{reference}
}
\startdata
$>90$   & 2.30 $\pm$ 0.01\tablenotemark{b} & \citet{Gulick2021WL17} \\
40--48  & 2.47 $\pm$ 2.33\tablenotemark{c} & Ep.2 \\
29--37  & 1.86 $\pm$ 0.22\tablenotemark{b} & Ep.1 \\
18--26  & 1.44 $\pm$ 0.12\tablenotemark{b} & Ep.3 \\
14--18  & 0.05 $\pm$ 0.58\tablenotemark{c} & Ep.4 \\
14--18  & 0.98 $\pm$ 0.57\tablenotemark{c} & Ep.5 \\
$<14$   & 0.34 $\pm$ 0.38\tablenotemark{b} & Ep.1 \\
$<14$   & 0.72 $\pm$ 0.36\tablenotemark{b} & Ep.2 \\
$<14$   & 0.10 $\pm$ 0.51\tablenotemark{b} & Ep.3 \\
$<14$   & 0.76 $\pm$ 0.32\tablenotemark{b} & Ep.4 \\
$<14$   & 1.41 $\pm$ 0.15\tablenotemark{b} & Ep.5 \\
\hline
\enddata
\tablenotetext{a}{Detailed observing date is provided in Table \ref{tab:obs}.} \vspace{-0.2cm}
\tablenotetext{b}{The error of spectral index is derived by the {\tt GNUPLOT fit} function.} \vspace{-0.2cm}
\tablenotetext{c}{As the number of available data points is only two, the error of spectral index is calculated by a standard error propagation: $1/{\rm ln (\nu_1/\nu_2)}\cdot\sqrt{\sigma_1^2/{\rm Flux_1^2}+\sigma_2^2/{\rm FLUX_2^2}}$.} \vspace{-0.2cm}
\end{deluxetable}

\section{Constraints on Maximum Grain Sizes}\label{sec:amax}
\subsection{Qualitative interpretation for the SED}\label{sub:sed}

The $\alpha_{\rm 100\mathchar`-345GHz}=2.30\pm0.01$ result at 100--345 GHz \citep{Gulick2021WL17} can be explained by marginally optically thick dust thermal emission (\citealt{Hildebrand1983}).
A plausible interpretation for the 14--48 GHz spectral feature is optically thick dust emission. Although the thermal dust emission at lower frequencies of 14--48 GHz is generally optically thinner, if localized high-density structures are present, the dust emission could be optically thick. The flux density of the optically thinner part decreases with wavelength very rapidly. Consequently, at lower frequencies, the emission from the optically thicker substructures can be seen.

A simple model that explains the $>$14 GHz SED of the WL~17 ring thus must include at least two distinct dust emission component: one low column density disk halo that is not optically thick at $\sim$200--345 GHz, and one high column density component that is optically thick even at $\lesssim$30 GHz.  
The optically thinner component has been probed in the spatially resolved spectral index analysis of \citet{Han2023WL17}.
If these two dust emission components have different spatial distributions (e.g., due to dust segregation), the observations at $>$200 GHz and $<$100 GHz may resolve different morphology.
This is consistent with what was found in the previous high angular resolution ALMA Band 3 and 7 observations \citep{Han2023WL17}.

An alternative possibility to explain the low spectral indices at 14--26 GHz is considering the emission of spinning nanometer-sized dust, which might have been detected in the planet-forming disk PDS~70 (\citealt{Liu2024ApJ...972..163L}) and a few other protoplanetary disks (\citealt{Greaves2018NatAs...2..662G,Hoang2018ApJ...862..116H,Greaves2022a})\footnote{Note that the GBT measurements of flux densities presented in \citet{Greaves2018NatAs...2..662G} and \citet{Greaves2022a}, in many cases, deviated largely from the JVLA measurements \citep{Chung2025Taurus}.}
In our present case study of WL~17, the emission of spinning nanometer-sized dust is not considered since 
it relies on substantial fine-tuning of the parameters in the radiative transfer modeling (c.f. Section \ref{sub:model}) to match the observed SEDs.
In addition, both the theories and the previous claimed detections in the protoplanetary disks remain uncertain.
If the emission of spinning nanometer-sized dust is indeed prominent at $\sim$20--30 GHz, a radiative transfer model without considering this emission mechanism  (Section \ref{sub:model}) will overestimate $a_{\mbox{\scriptsize max}}$ and dust masses ($M_{\mbox{\scriptsize dust}}$).

The $\alpha<$2 results and the time variations of flux densities and spectral indices at $<$14 GHz can be interpreted by time-varying free-free emission, which is moderately common in the Class 0/I stage (e.g., \citealt{Dzib2013ApJ...775...63D,Liu2014ApJ...780..155L,Coutens2019A&A...631A..58C}).
We do not consider synchrotron emission in our present interpretation since in sources that present bright free-free emission, synchrotron emission has a high chance to be obscured by free electron (\citealt{Gudel2002ARA&A..40..217G}).

We quantified our interpretation with a simplified radiative transfer model, which is detailed in the following Section \ref{sub:model}.

\subsection{SED modeling}\label{sub:model}

\subsubsection{Formulation}
We follow the approach of \citet{Liu2019c,Liu2021a} to model the invariant dust thermal emission and free-free emission, using the following formulation:

\begin{equation}\label{eqn:multicomponent}
    F_{\nu} = \sum\limits_{i} F_{\nu}^{i} e^{-\sum\limits_{j}\tau^{i,j}_{\nu}}, 
\end{equation}
where $F_{\nu}^{i}$ is the flux density of the dust or free-free emission component $i$, and $\tau^{i,j}_{\nu}$ is the optical depth of the emission component $j$ that obscures the emission component $i$  (e.g., the free-free emission around the central protostar may or may not be obscured by the optically thick dust).
The details of how we produced the SEDs of dust and free-free emission are given in Appendix \ref{appendix:mechanism}.

In our modelings, we found that the simplest model to fit the observed SEDs (Figure \ref{fig:sed}) needs to include two dust emission components that have very different column densities (c.f. Section \ref{sub:sed}), one free-free emission component for January 27, and two free-free emission components for February 05--21.
We have tried various abstracted geometry by altering the $\tau^{i,j}_{\nu}$ terms (e.g., the free-free emission may or may not be obscured by the optically thick dust). 
In the end, we found that in order to explain the observed SEDs (Figure \ref{fig:sed}), it is not necessary to consider mutual obscuration of the emission components (i.e., $\tau^{i,j}_{\nu}=0$ for all $i$ and $j$).
This may be explained by the low inclination angle of 34 deg of WL~17 ring.
Details of the individuals of emission components are discussed in Section \ref{subsub:SEDresult}.

\begin{deluxetable}{ llllll }
\tablecaption{Model parameters for dust emission\label{tab:dustmodel}}
\tablewidth{700pt}
\tabletypesize{\scriptsize}
\tablehead{
\colhead{Component} &
\colhead{$T_{\rm dust}\tablenotemark{a}$} &
\colhead{$\Sigma_{\rm dust}$} &
\colhead{$\Omega_{\rm dust}$} &  
\colhead{$a_{\rm max}$} &
\colhead{$M_{\rm dust}$}\\
&
\colhead{(K)} &
\colhead{(g\,cm$^{-2}$)} &
\colhead{($10^{-2}$ arcsec$^2$)} & 
\colhead{(mm)} &
\colhead{($M_{\rm Jup}$)} \\
} 
\startdata
\multicolumn{6} {c} { (Initial condition\tablenotemark{b}) } \\
Grown & 35 & 20   & 4.0 & 3.06 & $\cdots$ \\
Small & 35 & 0.12 & 29  & 0.045\tablenotemark{c} &  $\cdots$ \\\hline
\multicolumn{6} {c} { (Best fits) } \\
Grown & 35 &     35$^{+28}_{-20}$   & 4.1$^{+0.5}_{-0.4}$
& 4.2$^{+1.8}_{-2.1}$ & 3.1$^{+3.1}_{-1.9}$       \\
Small & 35 & 0.11$^{+0.19}_{-0.07}$ & 28.9$^{+45.9}_{-17.5}$ 
& 0.045\tablenotemark{c} & 0.069$^{+0.42}_{-0.059}$ \\
\enddata
\tablenotetext{a}{We artificially fixed dust temperature to 35 K, which is consistent with the temperature of 30--40 K at the ring used in previous studies \citep{Sheehan2017WL17,Gulick2021WL17,Han2023WL17}.}
\vspace{-0.2cm}
\tablenotetext{b}{The parameter range is set to vary by $\pm1$ order of magnitude from the initial value.}
\vspace{-0.2cm}
\tablenotetext{c}{We artificially fixed the $a_{\rm max}$ of the Small dust component to 45 $\mu$m.}
\end{deluxetable}

\subsubsection{Results of modeling}\label{subsub:SEDresult}

The best-fit parameters of our models are summarized in Table \ref{tab:dustmodel} and \ref{tab:freefreemodel}.
They were obtained using a routine that was based on the the Markov Chain Monte Carlo (MCMC) method, which was implemented using the Python-based package {\tt emcee} (\citealt{foreman-mackey+2013}).
We used 250 MCMC walkers, of which the initial positions were found by interactively searching the parameters that can fit the observational data approximately, which are provided in Tables \ref{tab:dustmodel} and \ref{tab:freefreemodel}. 
We then iterated each walker with 50,000 steps, allowing its position to be deviated from the initial position by $\pm1$ order of magnitude.
In these iterations, we employed Gaussian likelihood function and a flat prior function.
In the end, we discarded the initial 10,000 steps as burn-in steps, and determined the best-fit and 1-$\sigma$ uncertainty of each parameter based on the median and standard deviation of the MCMC samplers.
The detailed of the fittings and some caveats of the fitting results are elaborated in as follows. 

\paragraph{Dust emission components.}

The temperatures ($T_{\rm dust}$), column densities ($\Sigma_{\rm dust}$), maximum grain size ($a_{\rm max}$), and solid angle ($\Omega_{\rm dust}$) of the high and low column density dust emission components need to be described with eight free parameters. 
However, these free parameters are only constrained by the observations at three JVLA bands (K, Ka, and Q) and three ALMA bands (3, 6, and 7).
As a result, some (but not all) fitting parameters are degenerated.
The most obvious degeneracy is $T_{\rm dust}$ and $\Omega_{\rm dust}$, although the previous high angular resolution images have provided the upper limits for  $\Omega_{\rm dust}$ \citep{Sheehan2017WL17,Shoshi2024WL17}.
We artificially fix $T_{\rm dust}$ of both dust components to 35 K which yield reasonable $\Omega_{\rm dust}$ values. The $T_{\rm dust}$ of 35 K is consistent with the temperature of 30--40 K at the ring used in previous studies \citep{Sheehan2017WL17,Gulick2021WL17,Han2023WL17}.
The choice of $T_{\rm dust}$ does not have a large effect on the resulting $\Sigma_{\rm dust}$ and $a_{\rm max}$ values.
However, through the determination of $\Omega_{\rm dust}$, it will affect the estimates of $M_{\rm dust}$.
The chosen $T_{\rm dust}$ is not too high and thus the $M_{\rm dust}$ in Table \ref{tab:dustmodel} may be regarded as upper limits. 

The properties of the low column density component (hereafter the disk halo) are mainly constrained by the ALMA observations at $>$200 GHz, which included only two independent measurements. 
Besides fixing its $T_{\rm dust}$ to 35 K, we further artificially fix its $a_{\rm max}$ to 45 $\mu m$, which is comparable to the constraints given by the previous dust polarization observations on the other Class II disks (e.g., \citealt{Kataoka2016a,Kataoka2016b,Bacciotti2018,Hull2018ApJ...860...82H,Ohashi2019ApJ...886..103O,Ohashi2020}).
Under these assumptions, the values of $\Sigma_{\rm dust}$ and $\Omega_{\rm dust}$ were constrained. 
As the DSHARP dust absorption opacities at $\gtrsim$230 GHz are not sensitive to the assumed $a_{\rm max}$ values as long as $a_{\rm max}$ is smaller than 100 $\mu$m (c.f., \citealt{Liu2021ApJ...914...25L}), we may regard the $\Sigma_{\rm dust}$ and $M_{\rm dust}$ as fair order of magnitude estimates. 

The properties of the high column density component are constrained by the JVLA K, Ka, and Q band observations (18--48 GHz) and the ALMA Band 3 observations at 100 GHz \citep{Sheehan2017WL17}.
The $\sim$2 values of $\alpha$ is consistent with an optically thick dust slab.
The frequency variation of $\alpha$ at 18--48 GHz is consistent with an optically thick dust slab with $a_{\rm max}=4.2^{+1.8}_{-2.1}$ mm (Figure \ref{fig:sed}; Table \ref{tab:dustmodel}).
Both the $a_{\rm max}$ and $\Sigma_{\rm dust}$ should be regarded upper limits, as emission of spinning nanometer-sized dust (\citealt{Greaves2018NatAs...2..662G,Hoang2018ApJ...862..116H}) and optically thick free-free emission\footnote{In principle, the spectral index of optically thick free-free emission can be as high as 2.0. However, a realistic ionized wind around the protostar includes optically thin and optically thick structures. In this case, the highest achievable spectral index may be $\sim$0.6 \citep{Reynolds1986ApJ...304..713R}.} may potentially contribute to a higher fraction of 18--48 GHz emission than what has been considered in our current model.
Nevertheless, we argue that $a_{\rm max}$ should not be too much lower than 1 mm, otherwise, it is hard to make dust emission optically thick at Q band (40--48 GHz).
The high column density dust component may be illustrated as substructures that harbor grown dust (e.g., narrow rings or vortices), which can be discerned in future $\sim$30--50 GHz observations with better angular resolutions (e.g., $\lesssim$50 mas).

\paragraph{Free-free emission component.}

In our five epochs of observations, the flux densities at $<$8 GHz can be explained by a free-free emission source (hereafter free-free-I) that has low {\it EM} and high $\Omega_{\rm ff}$ values (Table \ref{tab:freefreemodel}).
In Table \ref{tab:freefreemodel}, the {\it EM} of free-free-I should be regarded upper limits, as the turnover frequency of this component was not resolved by our observations (Figure \ref{fig:sed}).
Nevertheless, {\it EM} cannot be arbitrarily low, which may make the values of $\Omega_{\rm ff}$ too large to be consistent with a point-like source in our observations (Section \ref{sec:obs}).

On February 05--21, the spectral indices are larger than 0 at $\gtrsim$10 GHz, which may be consistent with another optically thick, time variable free-free emission component (hereafter free-free-II).
The $\gtrsim$10 GHz turnover frequency of free-free-II infers the high {\it EM} values (Table \ref{tab:freefreemodel}).
In this case, for the flux densities to be lower than what were detected in our 2--20 GHz observations, the values of $\Omega_{\rm ff}$ needs to be low.
We found that the solid angle of free-free-II is at least 2 orders of magnitude smaller than that of free-free-I (Table \ref{tab:freefreemodel}).
Therefore, free-free-II  may be illustrated as a spatially unresolved ionized gas knot. 
Presently, our simplified model considered free-free-I and free-free-II as independent knots of ionized gas.
More complicated models may be probable, which is beyond the scope of our present study. 

As discussed in \S \ref{sec:discussion}, the WL 17's ring might be gravitationally unstable. 
Consequently, accretion bursts may occur and trigger an additional radio jet knot responsible for another free-free emission component (c.f. \citealt{Anglada1998Jet}) although there can also be other possibilities.

\begin{deluxetable}{l  cll }
\tablecaption{Model parameters for the free-free emission in 2024\label{tab:freefreemodel}}
\tablehead{
\colhead{Date} &
\colhead{$T_{e}$\tablenotemark{a}} &
\colhead{EM} &  
\colhead{$\Omega_{\rm ff}$} \\
\colhead{(UTC)} &
\colhead{(10$^{3}$ K)} &
\colhead{(10$^{7}$ cm$^{-6}$pc)} & 
\colhead{(10$^{-14}$sr) } \\
} 
\startdata
\multicolumn{4}{c}{free-free-I} \\
\multicolumn{4} {c} { (Initial condition\tablenotemark{b}) } \\
Jan. 27 &
8 &
0.10 &
70   \\
Feb. 05 &
8 &
0.1 &
40   \\
Feb. 18 &
8 &
0.2 &
25   \\
Feb. 20 &
8 &
0.2 &
30   \\
Feb. 21 &
8 &
0.2 &
15   \\
\multicolumn{4} {c} { (Best fits) } \\
Jan. 27 &
8 &
0.10$^{+0.14}_{-0.057}$ &
62$^{+77}_{-35}$   \\
Feb. 05 &
8 &
0.098$^{+0.13}_{-0.057}$ &
40$^{+55}_{-23}$   \\
Feb. 18 &
8 &
0.20$^{+0.22}_{-0.11}$ &
22$^{+26}_{-12}$   \\
Feb. 20 &
8 &
0.19$^{+0.19}_{-0.092}$ &
30$^{+27}_{-15}$   \\
Feb. 21 &
8 &
0.21$^{+0.34}_{-0.13}$ &
12$^{+19}_{-7.4}$   \\\hline
\multicolumn{4}{c}{free-free-II} \\
\multicolumn{4} {c} { (Initial condition\tablenotemark{b}) } \\
Feb. 05 &
8 &
60 &
0.15   \\
Feb. 18 &
8 &
50 &
0.075   \\
Feb. 20 &
8 &
70 &
0.09   \\
Feb. 21 &
8 &
100 &
0.15   \\
\multicolumn{4} {c} { (Best fits) } \\
Feb. 05 &
8 &
58$^{+160}_{-42}$ &
0.081$^{+0.16}_{-0.053}$   \\
Feb. 18 &
8 &
110$^{+170}_{-84}$ &
0.063$^{+0.11}_{-0.038}$   \\
Feb. 20 &
8 &
79$^{+120}_{-54}$ &
0.089$^{+0.10}_{-0.059}$   \\
Feb. 21 &
8 &
91$^{+110}_{-48}$ &
0.14$^{+0.10}_{-0.072}$   \\
\enddata
\tablenotetext{a}{We artificially fixed dust temperature to 8000 K.}
\vspace{-0.2cm}
\tablenotetext{b}{The parameter range is set to vary by $\pm1$ order of magnitude from the initial value.}
\end{deluxetable}

\section{Discussion} \label{sec:discussion}

In this section, based on the results from our SED modeling in \S \ref{sub:model}, we discuss the gravitational stability of WL 17's ring and the potential planetary core mass that could form through pebble accretion within the structure.

\subsection{Gravitational stability} \label{subsec:GI}

Assuming a gas-to-dust mass ratio of the canonical value of 100 \citep{bohlin78}, the total disk mass (gas~$+$~dust) of WL 17 is approximately 0.3 $M_\sun$ (Table~\ref{tab:dustmodel}), while WL 17 itself could have a mass of 0.3~$M_\sun$ \citep{Han2023WL17}. A large disk-to-star mass ratio exceeding 0.1 can potentially lead to gravitational instability \citep[e.g.,][]{kratter16}. However, as discussed in \S \ref{sub:sed}, our dust mass estimate likely represents an upper limit. Spatially resolved observations could provide more accurate constraints on the dust mass.

In gravitationally unstable disks, spiral arms are expected to be present in the gas disk \citep[e.g.,][]{dong15gi}. Additionally, a part of spiral arms can be identified as azimuthal asymmetries in the dust disk \citep[e.g.,][]{dong2018mwc758}. 
Indeed, WL 17 might exhibit such azimuthal asymmetries in its dust disk \citep{Sheehan2017WL17,Gulick2021WL17,Han2023WL17,Shoshi2024WL17}. However,
estimating the Q-value comes with a major uncertainty, namely the gas-to-dust mass ratio. If the ring represents a dust trap, the local gas-to-dust ratio could be significantly lower than 100. To further constrain on the gas mass in the WL 17 system, observations of gas-mass tracers such as HD \citep{Bergin2013HD} and/or analyses of line pressure broadening \citep{Yoshida2022Broadening} would provide valuable insights.

\subsection{Potential planetary core mass by pebble accretion} \label{subsec:pebble}

Pebble accretion has recently gained attention due to its potential to significantly accelerate the planet core growth \citep{Ormel2010Pebble, Lambrechts2012Pebble}. Even at large distances from the central star (e.g., at $>50$ au), planetesimals can grow into the cores with masses exceeding a range of approximately 5--10 $M_\earth$ within 1 Myr \citep{Jang2022,Jiang2023Pebble}. This mass is referred to as the critical core mass, which is necessary to trigger runaway gas accretion to form gas giant planets \citep[e.g.,][]{Mizuno1980CoreMass}.

\citet{Jiang2023Pebble} provided a semi-analytical formula for the typical planet core formed in the ring where pebbles accumulate, described as

\footnotesize
\begin{equation}\label{eq:Mp_Mring}
\begin{aligned}
    M_{\rm core} &= 18 M_\earth \times
    \left(\frac{f_{\rm rs}}{1.8}\right)^{-\frac{3}{4}}
    \left(\frac{f_{\rm rp}}{1.4}\right)^{\frac{3}{2}}
    \left(\frac{f_{\rm mg}}{1}\right)^{-\frac{3}{4}}
    \left(\frac{M_{\rm dust}}{30 M_\oplus}\right)^{\frac{3}{4}}
    \\
    &\times
    \left(\frac{\Sigma_{\rm g}r_0^2}{0.002M_\odot}\right)^{-\frac{3}{4}}
    \left(\frac{\rm St}{0.01}\right)^{\frac{1}{2}}
    \left(\frac{h_{\rm g}}{0.07}\right)^{\frac{3}{2}}
    \left(\frac{M_\star}{M_\odot}\right),
\end{aligned}
\end{equation}
\normalsize
where $f_{\rm rs}=1.8$ is a numerical pre-factor, $f_{\rm rp}=1.4$ is a fit constant, $f_{\rm mg}=1$ is a type-I migration pre-factor \citep{Jiang2023Pebble}, $M_{\rm dust}=1007M_\oplus$ is the dust mass in the WL 17's ring (Table \ref{tab:dustmodel}), $\Sigma_{\rm g}$ is the gas surface density in ring, $r_0=16$ au is the ring location \citep{Sheehan2017WL17,Shoshi2024WL17}, $St$ is a Stoke number, $h_{\rm g}=0.084$ is a ring aspect ratio of gas\footnote{The gas scale height $h$ is the ratio of the gas sound speed to the angular velocity ($h=c_s/\Omega$), then the ring aspect ratio can be written as $h_{\rm g} = h/r \approx 0.03 (\frac{M_{\rm star}}{M_\sun})^{-\frac{1}{2}} (\frac{T_{\rm mid}}{300{\rm K}})^{\frac{1}{2}} (\frac{r}{1{\rm au}})^{\frac{1}{2}}$, where $T_{\rm mid}$ is a disk midplane temperature with 35~K (\S~\ref{subsub:SEDresult}).} at $r_0=16$ au, and $M_\star$ is $0.3 M_\sun$ \citep{Han2023WL17}, respectively.

The Stokes number is defined as  
\begin{eqnarray}
{\rm St}=\frac{a_{\rm max} \rho_{\rm s}}{\Sigma_{\rm g}}\frac{\pi}{2},
\end{eqnarray}
where $a_{\rm max}$ is 4.2 mm (Table \ref{tab:dustmodel}), $\rho_{\rm s}$ is a mean material density of the dust particle of 1.675 g/cm$^3$ for the DSHARP dust grains \citep{Birnstiel2018} and $\Sigma_{\rm g}$ is gas surface density, respectively. Assuming a gas-to-dust mass ratio of 10 and 100, the Stokes number is calculated to be $3.2\times10^{-3}$ and $3.2\times10^{-4}$, respectively.

Assuming a gas-to-dust mass ratio of 10 and 100, the planetary core mass calculated from equation (\ref{eq:Mp_Mring}) is approximately 16.4 and 1.0 $M_\earth$, respectively\footnote{The pebble isolation mass ($M_{\rm iso}$) is estimated as $M_{\rm iso}\approx25(h_{\rm g}/0.05)^3(M_{\star}/M_\sun)M_\earth\approx36M_\earth$ \citep[e.g.,][]{liu2019a}, which is larger than the expected planetary core mass of 1.0--16.4 $M_\earth$.}. These values are derived using the estimated dust mass in Table \ref{tab:dustmodel}. As noted in \S \ref{sub:sed}, our estimates of dust mass and maximum grain size likely represent upper limits. Accordingly, the calculated planetary core mass should also be considered an upper limit.

The critical core masses with a gas-to-dust mass ratio of 1, 10, and 100, for triggering runaway gas accretion, are expected to be approximately 12, 6, and 4 $M_\earth$, respectively \citep{Mizuno1980CoreMass}. Given that the expected planetary core mass exceeds approximately 6 $M_\earth$ in case of the gas-to-dust mass ratio of 10, it might indicate that a gas giant planet could potentially grow in the WL 17 ring. A less massive gas disk corresponds to the particles with a higher Stokes number. A more massive core could form. As discussed in \S \ref{subsec:GI}, further estimates of the disk mass are necessary. 

\section{Conclusion} \label{sec:conclusion}

We observed the WL 17 ring using the JVLA at centimeter wavelengths, ranging from $\lambda=0.6$ cm (48 GHz) to 15.0 cm (2 GHz). We successfully detected signals across all wavelengths, except at 15.0 cm. Based on the spectral indices, we attribute the shorter wavelength emission ($\lambda\lesssim3$ cm) to dust rather than free-free emissions. However, the WL 17 ring is not spatially resolved in our observations due to the angular resolution being greater than 0\farcs5.

The emission in the frequency range of 2--340 GHz was analyzed using a simple SED model that incorporates two dust components and two free-free emission components. It is important to note that the results of our modeling are influenced by the uncertain properties of the dust. The model includes dust ring components with maximum grain sizes ($a_{\rm max}$) of 4.2 mm and 0.045 mm, as well as free-free emission components with varying electron densities. Our analysis suggests that grain growth in the WL 17 ring has progressed to millimeter-sized grains.

If the gas-to-dust mass ratio is less than approximately 10, pebble accretion within the ring structure might facilitate the formation of a giant planet.  However further estimates of the disk mass and the maximum grain size are necessary.

While our SED analysis suggests the grain size of 4.2 mm in the WL 17 ring, the grain size could potentially be larger if the ring exhibits azimuthal asymmetries. Due to the large beam size of more than 0\farcs5 in our C-configuration observations, the ring structure is not spatially resolved (Figures \ref{fig:cband-image1}--\ref{fig:cband-image6}). If azimuthal asymmetries are present at longer wavelengths, thermal emissions from larger dust grains may be required to account for the observed flux density because emitting areas of azimuthal asymmetries are smaller. Future JVLA observations in A-configuration, with a spatial resolution of less than 0\farcs1, would provide valuable insights into this possibility.

\bigskip

The authors thank the anonymous referee for a constructive report.
The National Radio Astronomy Observatory is a facility of the National Science Foundation, operated under a cooperative agreement by Associated Universities, Inc.
This study was supported by JSPS KAKENHI Grant Number 23K03463 and 25K07376.
H.B.L. is supported by the National Science and Technology Council (NSTC) of Taiwan (Grant Nos. 111-2112-M-110-022-MY3, 113-2112-M-110-022-MY3).

\software{
          astropy \citep{Astropy2013,Astropy2018,Astropy2022},  
          Matplotlib \citep{Hunter2007Matplotlib},
          Numpy \citep{VanDerWalt2011}
          }

\facilities{JVLA, ALMA}


\bibliography{sample63}{}
\bibliographystyle{aasjournal}

\appendix

\section{Emission mechanisms}\label{appendix:mechanism}
\subsection{Dust emission}\label{sub:dust}

The spectral profiles of dust were evaluated based on  Equations (10)--(20) presented in \citet{Birnstiel2018}, which is the analytic solution of the equation of radiative transfer for a geometrically flat isothermal dust slab (c.f. \citealt{Miyake1993}), which adopted the  Eddington-Barbier approximation. 
The formulation takes into account dust emission/absorption and self-scattering self-consistently. 
Given the assumption of dust opacities (introduced below), the free parameters for each dust component are dust temperature ($T_{\rm dust}$), dust mass surface density ($\Sigma_{\rm dust}$), solid angle ($\Omega_{\rm dust}$), and $a_{\rm max}$.
We assumed that these free parameters do not vary with time (at least, during the observations we compared with).

We adopted the default DSHARP dust opacity model, which assumes a compact dust morphology (\citealt{Birnstiel2018}).
The composition includes  water ice \citep{Warren1984ApOpt..23.1206W}, astronomical silicates \citep{Draine2003ARA&A..41..241D}, troilite, and refractory organics (\citealt{Henning1996A&A...311..291H}).
As the DSHARP opacity table has been widely applied \citep[e.g.,][]{Hashimoto2022ZZTauIRS,Hashimoto2023AJ....166..186H,Liu2024ApJ...972..163L,Liu2024A&A...685A..18L,Aso2024TMC-1A,Riviere-Marichalar2024ABAur,Guerra-Alvarado2024IRAS4A1}, this choice of opacity table allows a straightforward comparison with several previous observational studies.
As the actual dust properties remain largely uncertain, we refer to the Appendix C of \citet{Chung2024ApJS..273...29C} and Section 5 of \citet{Guidi2022HD163296} for the comprehensive discussion about the effects of assuming different dust compositions and porosity in the SED modeling.

The sizes of dust grains in protoplanetary disks have a broad distribution (\citealt{Birnstiel2018}).
For simplicity, when evaluating the size-averaged opacities, we assumed a power-law grain size distribution function (i.e., $n(a)\propto a^{-q}$) in between the minimum and maximum grain sizes ($a_{\mbox{\scriptsize min}}$, $a_{\mbox{\scriptsize max}}$), where the values of $q$ and $a_{\mbox{\scriptsize min}}$ were assumed to be 3.5 and $10^{-4}$ mm, respectively.
The size-averaged opacities depend very weakly on $a_{\mbox{\scriptsize min}}$. 

\subsection{Free-free emission}\label{sub:freefree}

Free-free emission is the thermal emission of free electrons, which can be described by the electron temperature $T_{e}$ and the optical depth $\tau^{\rm ff}$.
The value of $\tau^{\rm ff}$ at frequency $\nu$ depends on $T_{e}$ and the emission measure $EM=\int n_{e}^{2}d\ell$, where $n_{e}$ is the electron number volume density.
To evaluate $\tau^{\rm ff}$, we adopted Equation (10) of \citet{Keto2003ApJ...599.1196K}, which is reproduced as follows:
\begin{equation}
\label{eq:tauff}
\tau_{\nu}^{\rm ff}=8.235\times10^{-2}\left(\frac{T_{e}}{\mbox{K}}\right)^{-1.35}\left(\frac{\nu}{\mbox{GHz}}\right)^{-2.1}\left(\frac{\mbox{EM}}{\mbox{pc\,cm$^{-6}$}}\right),
\end{equation}
With the formulation we adopted, the spectral indices can range from $-$0.1 (optically thin) to 2.0 (optically thick).
Flux density of free-free emission $F^{\rm ff}_{\rm \nu}$ was evaluated as  $F^{\rm ff}_{\rm \nu}=T_{e}(1-e^{-\tau_{\nu}^{\rm ff}})$

The temperatures of the thermal radio jets have not been very well constrained by observations.
The innermost part of the thermal radio jets may be as hot as $\gtrsim$10$^{4}$ K (c.f., \citealt{Reynolds1986ApJ...304..713R}).
In this work we assumed $T_{e}=$8000 K .
With this assumption and Equation (\ref{eq:tauff}), we can see that the turnover frequency where free-free emission is having a transition from the optically thin regime to optically thick regime is determined by {\it EM} (i.e., $n_{e}$).
Observing the frequency variations of the spectral indices of free-free emission therefore can constrain the electron density $n_{e}$ in the ionized gas source. 
We assumed different values of {\it EM} and $\Omega_{\rm ff}$ when fitting the observations taken at different dates to take time variability into consideration. 

\end{document}